# Low Velocity Granular Drag in Reduced Gravity

D. J. Costantino, J. Bartell, K. Scheidler, P. Schiffer*

*Department of Physics and Materials Research Institute, Pennsylvania State University, University Park, PA 16802, USA*

**Abstract**

We probe the dependence of the low velocity drag force in granular materials on the effective gravitational acceleration ($g_{eff}$) through studies of spherical granular materials saturated within fluids of varying density. We vary $g_{eff}$ by a factor of 20, and we find that the granular drag is proportional to $g_{eff}$, i.e., that the granular drag follows the expected relation $F_{probe} = \eta \rho_{grain} g_{eff} d_{probe} h_{probe}^2$ for the drag force, $F_{probe}$ on a vertical cylinder with depth of insertion, $h_{probe}$, diameter $d_{probe}$, moving through grains of density $\rho_{grain}$, and where $\eta$ is a dimensionless constant. This dimensionless constant shows no systematic variation over four orders of magnitude in effective grain weight, demonstrating that the relation holds over that entire range to within the precision of our data.

Granular materials, collections of classical particles that interact only through contact forces, display a wide range of complex properties that emerge from their collective interactions [1]. Among the most intriguing granular phenomena are those that result when the grains "jam" locally to resist an externally applied pressure or force, creating a skeleton of connected grains that provide structural strength against a distortion of the grain pack [2]. One result of such jamming on a local scale is the drag force resisting the low-velocity motion of an object through a granular sample [3,4,5,6,7,8,9,10,11,12]. This granular drag is unrelated to the surface friction between the object and the grains [8], but rather arises from the necessary dilation and local rearrangement of the jammed grains, typically allowed by a free top surface. The granular drag force is thus nearly velocity-independent in the low velocity regime, since it is not associated with the transfer of momentum. Previous studies of the drag on intruders moving through a granular sample have both directly measured the resulting drag force [3-12] and also used simulations [13] as well as imaging of the grains [7,11,14]. Furthermore, in two dimensional systems, imaging of the grains can even access the inter-grain forces resulting from the drag process [7].

Gravity plays an essential role in all reported measurements of granular jamming and drag in three dimensions, since the granular packs are held in place by the gravitational force even when the top surface of the pack is free. In the present work, we explore the dependence of the drag force on the effective gravitational acceleration, $g_{eff}$, by immersing the grains in fluids of different densities. Over a broad range of gravitational forces on the grains, we find that the granular drag force is proportional to $g_{eff}$ to within the precision of our measurements, in agreement with expectations for the low velocity regime.

Our apparatus, shown schematically in figure 1, is designed to measure the granular drag force on a vertical cylinder moving horizontally through a granular bed (details are available in [15]). Following the method used previously by our group [4,8], the grains were contained in a 16 cm diameter cylindrical glass bucket rotating slowly around a vertical axis. A stainless steel cylinder was mounted vertically within the grains and supported on a bearing that allowed free rotation around the same vertical axis. As the bucket rotated, the grains carried the cylinder around the bearing until it hit a fixed stop. We integrated a force cell within the stop, allowing us to measure the force required to prevent the cylinder from moving with the grains, thus measuring the granular drag force.

Our granular samples consisted of glass spheres ($d_{grain}$ = 0.048 ± 0.04 cm, $\rho_{grain}$ = 2550 kg / m$^3$) [16] or polystyrene spheres ($d_{grain}$ = 0.096 ± 0.009 cm, $\rho_{grain}$ = 1050 kg / m$^3$) [17] although almost all data reported below are for glass grains. The cylindrical probes could be varied in diameter, $d_{probe}$, and depth of insertion, $h_{probe}$, as discussed below. We took data while the bucket rotated at 4.5 – 5.5 mHz, corresponding to the probe having a speed of ~ 1.1 mm/s relative to the grains or about two grain diameters per second for the glass grains. Before beginning an experimental run, we rotated the bucket approximately 20 times faster than the measurement velocity for at least five revolutions before reducing the speed and taking data, in order to remove possible internal structures within the grain pack (data taken without stirring displayed considerably higher scatter). There was a delay on the order of 10 minutes or less between the end of the stirring and the start of data taking as the platform's speed was adjusted. We took data for ~ 5 rotations and averaged the results, and then this average value was averaged for at least three independent data runs to determine the drag force, $F_{probe}$; error bars in the plotted data correspond to the standard deviation among the different runs. We filled the buckets to 15-17 cm deep and took data for $h_{probe}$ up to 14 cm, with ~2 cm or more between the bottom of the cylinder and the bucket bottom. The grain packing was measured to be within the range of 61±2% for all samples, with the uncertainty arising from the slightly uneven surface of the grains. The data were unaffected by variation of the radial position of the cylinders within the bucket, which were 3.5 cm from the center of the beaker for all of the data below, and showed a very weak dependence on the speed of the probe, ~ 0.2 N/(m/s), consistent with previous studies [18]. As the probe moved through the pile, it created a bulge of grains on the surface in front of it, presumably associated with the dilation of grains that accompanied the reorganization of the jammed region in front of the probe.

Our group has previously reported that the behavior of $F_{probe}$ in the low velocity regime is consistent with simple mean-field expectations for quasi-static behavior, and can be expressed as

$$F_{probe} = \eta \rho_{grain} g d_{probe} h_{probe}^2 \qquad [1]$$

where $g$ is the acceleration due to gravity and $\eta$ is a dimensionless parameter. In these previous studies the dependences on $g$ and $\rho_{grain}$ were simply assumed for dimensional reasons and not tested [4,8], although a recent fluidized bed study showed that the granular drag disappears at the point of fluidization [12]. To test those dependences, we saturated our glass grains under liquids of varying densities matching fractions of the density of glass. By doing so, the liquid buoyant force effectively reduced the acceleration due to gravity, resulting in an effective gravitational acceleration of $g_{eff} = g\left(1 - \frac{\rho_{liquid}}{\rho_{grain}}\right)$. Our saturating fluids were air, denatured ethanol ($\rho_{liquid}$ = 800 kg/m$^3$), water ($\rho_{liquid}$ = 1000 kg/m$^3$), and water solutions of lithium heteropolytungstate (LST) [19]. LST has a high density, and thus by making several different concentration LST solutions, we were able to increase the fluid density to a range of values between the density of water and the density of glass. Because we are primarily interested in the buoyant effects of the LST solution, we labeled our LST solutions as *n*% LST, where *n* is the density of the solution expressed as a percentage of the density of glass. Our maximum density solution was ~95% LST, giving us a range in the effective gravitational acceleration from $g_{eff}$ = 9.8 m/s$^2$ in air down to $g_{eff}$ = 0.46 m/s$^2$ in ~95% LST. Note that we chose fluid densities that were less than the density of glass, so that the grains were resting on the bottom of the container and the packing fraction was constant for all samples. The grains used for the air measurements had previously been submerged in water and ethanol and were then dried; grains that had not been submerged yielded a ~20-30% lower drag force, presumably associated with altered inter-particle properties due to the presence of fines (microscopic particles) that could roll between the grains; they also generated inconsistent results at comparable depths within piles of different sizes. Note that the level of the liquid was always kept ~ 1–3 cm above the top of the grains, so that our data were not affected by capillary forces [20].

Since we are introducing a liquid to the interstitial space between the grains, we must consider the possible effects of viscous forces on both the cylinder and the grains. The maximum viscosity of our fluids was for the highest density LST and was < 5 mPa s [19]. The calculated fluid drag on the largest cylinder would therefore be < 10 µN at our velocities and thus negligible compared to the granular drag [21]. The possible effects of viscous drag on the grains are potentially more important. Assuming simple Stokes drag, as is appropriate for our low Reynolds number, the terminal speed of a grain in free fall is ~3 mm/s for the fluid with the highest density and viscosity (~95% LST). For the extreme case of a grain travelling at the speed of the cylinder, the viscous drag on a grain is thus approximately a third its apparent weight ($W_{app} = m_{grain}g_{eff}$), although for all other liquids the drag is considerably smaller (i.e., $< 0.2 W_{app}$ for 90% LST and $< 0.1 W_{app}$ for other fluids). The lower effective weight combined with the viscous force will slow the grain dynamics. In the cases of the smallest $W_{app}$, this effect could be large enough that the system would no longer be considered quasi-static and thus have a different dilation of the grains associated with motion of the probe through the grains. On the other hand, due to the complex non-linear behavior of the grain motion and the absence of a grain-scale probe in our experiments, we cannot ascertain directly if this is the case. As evidenced below, however, we find that the measured drag force follows the simple predication of equation 1, suggesting that viscous effects on grain motion had little impact on the drag experienced by the cylinders. Similarly, the data suggest that liquid lubrication of the grain-grain contacts had little impact on the results.

To investigate the formula for $F_{probe}$ found by [4,8], we plot $F_{probe}$ vs. $h_{probe}$ for several fluids in Figure 2. Since we expect $F_{probe}(h_{probe})$ to be quadratic, we have fit each set of points to function of the form $F_{probe} = a\, h_{probe}^{b}$. For $d_{probe} = 0.635$ cm and $d_{probe} = 1.27$ cm, we find the average value of $b$ for all fluids tested to be $2.3 \pm 0.3$ and $2.4 \pm 0.3$, respectively, thus verifying the quadratic dependence on $h_{probe}$. The values of the exponent $b$ above 2.0 may be due to slight variations in the grain packing with depth, although we are unable to test for such variations, and they should not qualitatively affect the results below. We test the dependence on $d_{probe}$ in our data by assuming the quadratic dependence on depth of insertion -- since each set of data were taken at slightly different depths in different fluids, we rescale the measured force for comparison and plot ($F_{probe}$ / $h_{probe}^2$) vs. $d_{probe}$ in Figure 3. The lines through the origin in Figure 3 clearly demonstrate the linear dependence of the drag force on the cylinder diameter in the different fluids. As evidenced in both Figure 2 and Figure 3, $F_{probe}$ depends strongly upon $g_{eff}$, and we plot this dependence explicitly in Figure 4, holding $d_{probe}$ and $h_{probe}$ constant at different values. As can be easily seen in the figure, $F_{probe}$ is proportional to $g_{eff}$ as expected, although there is a slight negative intercept to the linear fits shown (of order 0.2 N with an uncertainty of similar magnitude).

To provide a further test of the dependence of $F_{probe}$ on $g_{eff}$, we combine the data from all of our measurements for each value of $g_{eff}$. To do so, we take the measured values of $F_{probe}$ for each set of conditions, and we reframe Equation 1 to calculate $\eta = F_{probe} / \left( \rho_{grain} g_{eff} d_{probe} h_{probe}^2 \right)$. We then average $\eta$ over all measured values of $F_{probe}$ for each value of $g_{eff}$. The results are plotted against $g_{eff}$ in the main panel of Figure 5, where we see that $\eta$ appears to be constant (2.7 ± 0.4) over the full factor of 20 variation in $g_{eff}$ (although, given the scatter in the data, we recognize that a more precise characterization could reveal a small variation with $g_{eff}$). The apparently constant value of $\eta$ strongly suggests that the lubrication and viscous effects of the liquids did not alter the grain dynamics in a way that affected the drag force. Furthermore, the data for our plastic spheres (measured in air only) are fully consistent with the other data, suggesting that the value of $\eta$ is generic to drag in spherical grains, and is not specific to the surface properties or density of the glass spheres used for varying $g_{eff}$. In the inset to Figure 5, we extend the results by comparing data from our previous study of the drag force in glass spheres of varying diameter, ranging up to 5 mm [8]. To include the data from that study, we plot $\eta$ as a function of the apparent weight of the grains ($W_{app}$), and we find that $\eta$ is constant to within the scatter of the data over a span of more than four orders of magnitude in $W_{app}$. The results provide strong support for Equation 1 and its intrinsic dependence on the gravitational force, which had not been examined in previous granular drag studies.

Our results provide a window into the properties of granular materials in a reduced gravity environment, and they strongly support the framing of Equation 1, and the proportionality of the drag force to $g_{eff}$. The exact value of $\eta$ (presumably different for non-spherical grains) may be related to the volume of grains that is perturbed by the drag process, since the dilation of the grains is an important factor in the granular drag process [11]. Indeed, the reduction in granular drag with reduced gravity could be attributed to subtle changes in granular dilation with density matching, a factor to which our measurements are not sensitive. In addition to the implications for granular drag, by demonstrating the possibilities of reducing effective gravity while leaving grain properties otherwise essentially unchanged, our results open a range of possibilities for further studies of three-dimensional granular materials in reduced gravity conditions. The behavior of granular materials in reduced gravity environments should be of direct relevance to potential grain processing activity in earth orbiting satellites, as well as future mining operations on the surfaces of asteroids. Studies of three dimensional force chains and other static properties of grain packs could be especially revealing when the gravitational force becomes a tunable quantity, rather than a fixed constant of the system, since those properties are intrinsically dependent on gravitational force to hold the grains together except in the rare cases when they are fully confined.

We acknowledge support from NASA grant NAG3-2384 and the NSF REU program. We thank Sid Nagel for his helpful conversation and suggestions.

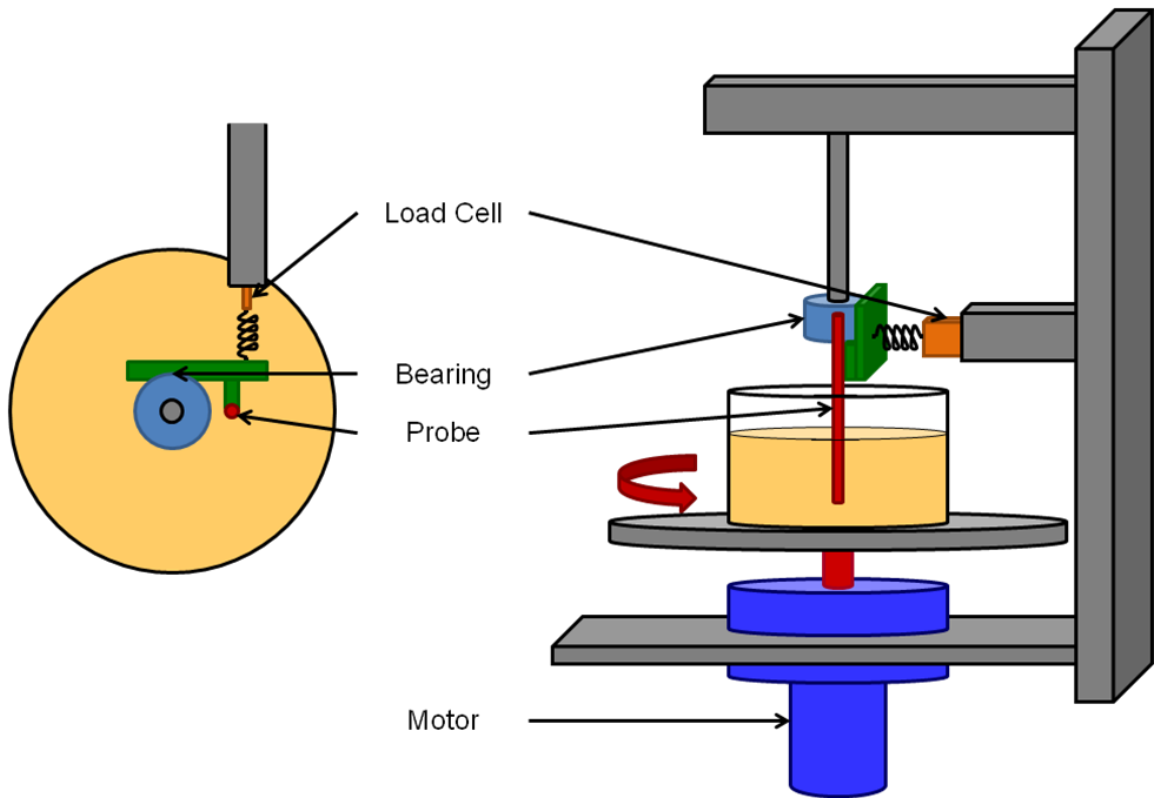

Figure 1: (Color online) Experimental apparatus as described in the text. The image on the left is a top view, while the image on the right is a side view that also shows the support structure of the experiment.

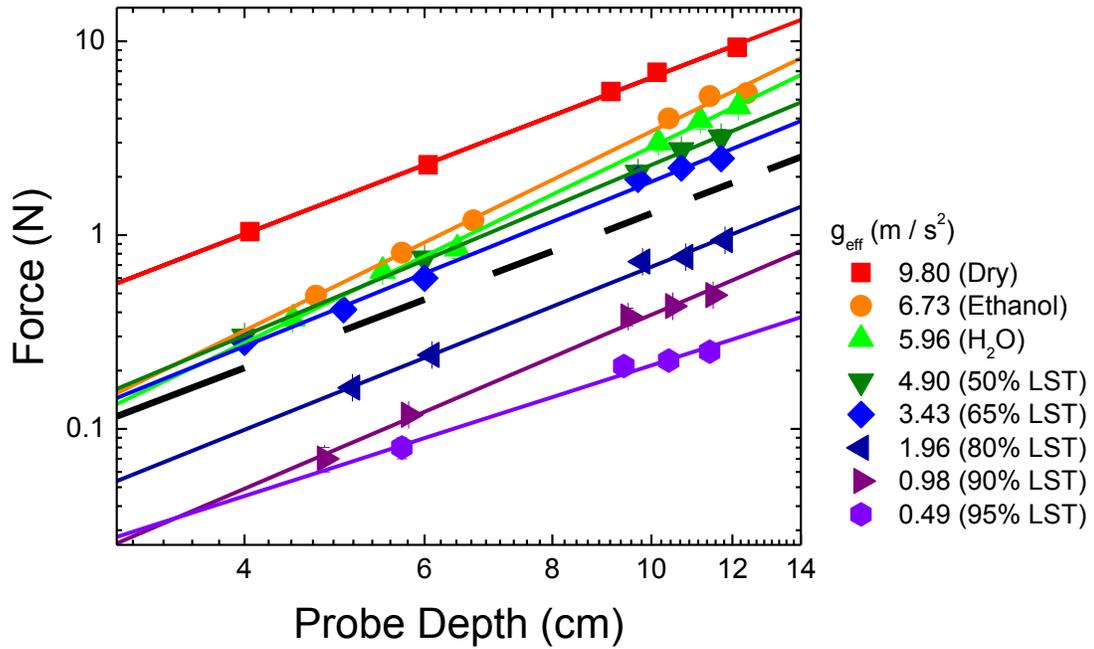

Figure 2: (Color online) $F_{probe}$ vs. $h_{probe}$ for all fluids for $d_{probe} \sim 0.635$ cm. The solid lines are fits of the form $F_{probe} = ah_{probe}^{b}$ as described in the text. The dashed line shows quadratic behavior for comparison.
11

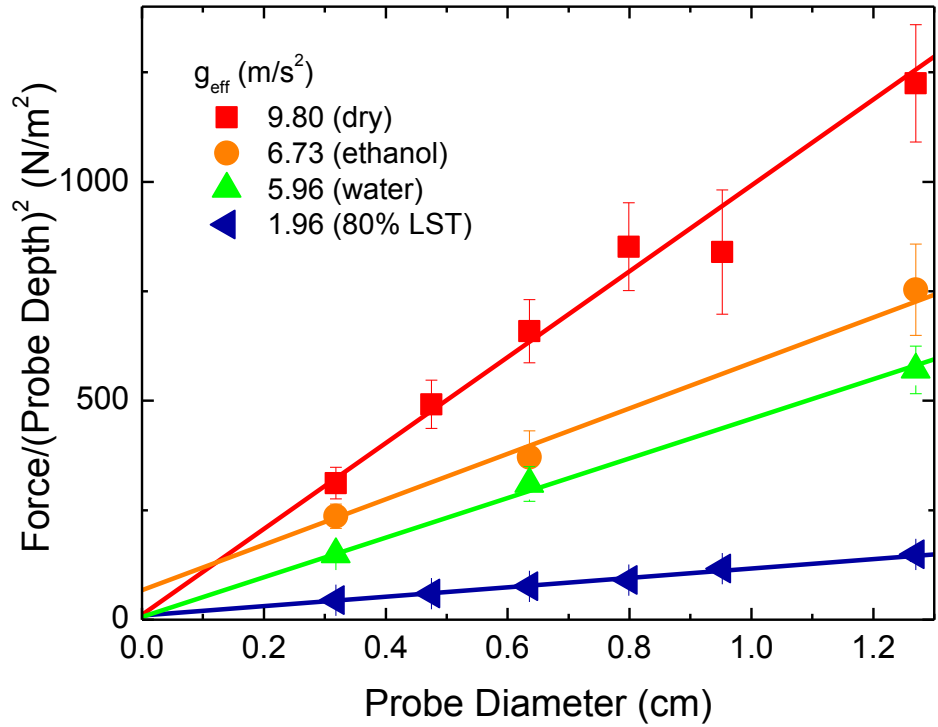

Figure 3: (Color online) The linear depth dependence of the drag force demonstrated by ($F_{probe}/h_{probe}^2$) vs. $d_{probe}$, where the data are scaled to account for different values of $h_{probe}$; the lines are linear fits to the data.

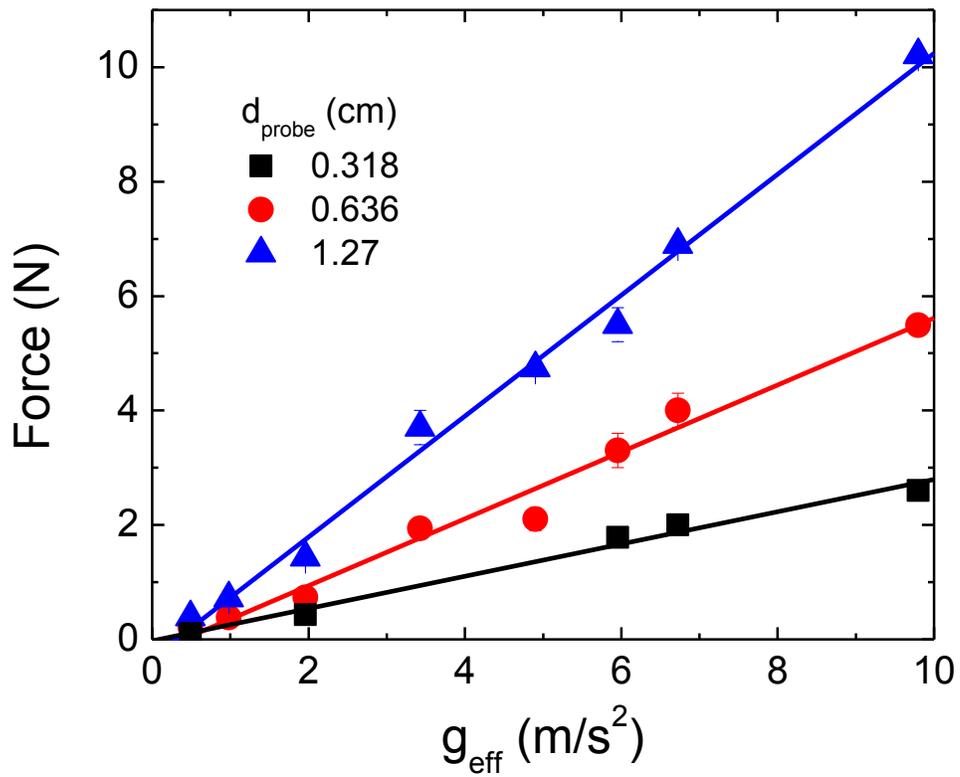

Figure 4: (Color online) The linear dependence of $F_{probe}$ on $g_{eff}$, shown for three different diameter probes at $h_{probe}$ ~ 10 cm; the lines are linear fits to the data.

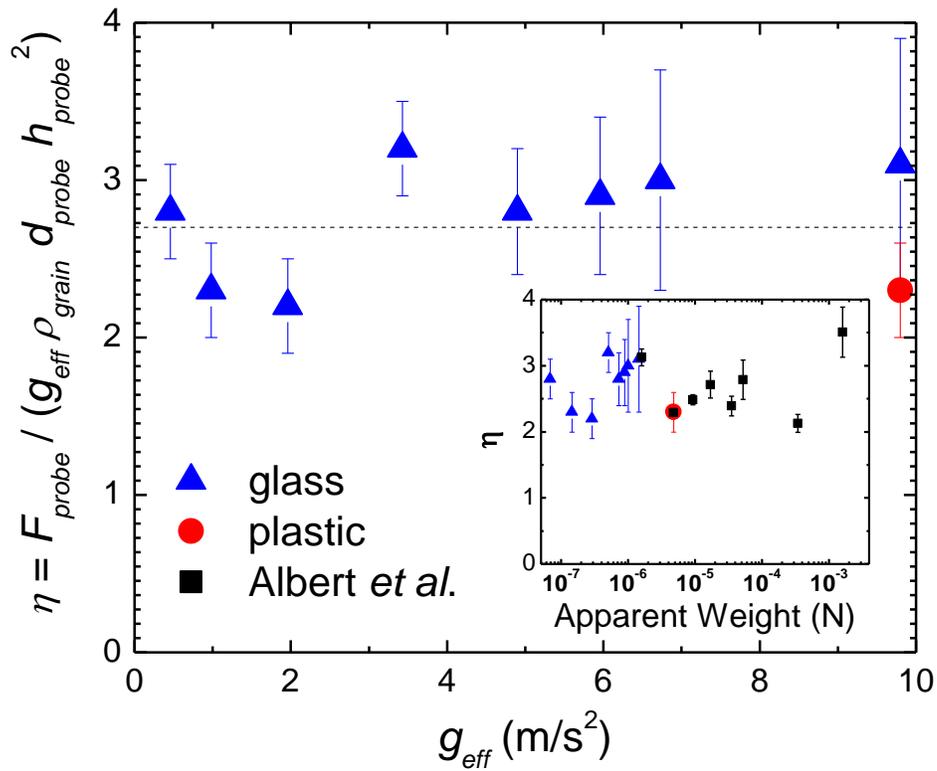

Figure 5: (Color online) The unitless drag coefficient $\eta$ vs. $g_{eff}$. Inset: $\eta$ vs. the apparent weight, $W_{app}$, for our data as well as data from [8] (different diameter glass spheres in air).